\documentstyle[11pt]{article}

\newcommand{\la}[1]{\label{#1}}

\renewcommand{\lg}{\langle}
\newcommand{\rg}{\rangle}

\newlength{\numlen}
\newcommand{\n}{\settowidth{\numlen}{0}\makebox[\numlen]{}}
\newcommand{\cen}[1]{\multicolumn{1}{c}{#1}}

\newlength{\indeksinpituus}

\newlength{\ypit}

\newcommand{\be}{\begin{equation}}
\newcommand{\ee}{\end{equation}}
\newcommand{\ba}{\begin{eqnarray}}
\newcommand{\ea}{\end{eqnarray}}

\newcommand{\etal}{{et al.\ }}
\newcommand{\eq}{eq.~}

\newcommand{\fig}{fig.~}

\newcommand{\nr}[1]{(\ref{#1})}

\newcommand{\h}{{\hspace{0.5 cm}}}

\newcommand{\den}{n}
\newlength{\poffset}
\begin{document}

\begin{titlepage}
\hfill CERN-TH.6654/92
\begin{centering}
\vfill

{\bf MULTICANONICAL CLUSTER ALGORITHM\\
 AND THE 2-D 7-STATE POTTS MODEL}

\vspace{1cm}
K. Rummukainen

\vspace{1cm}
{\em Theory Division, CERN\\ CH-1211 Geneva 23, Switzerland}

\vspace{1cm}
{\bf Abstract}

\end{centering}

\vspace{0.3cm}\noindent

I present a hybrid-like two-step algorithm, which combines a
microcanonical update of a spin system using demons, with a
multicanonical demon refresh.  The algorithm is free from the
supercritical slowing down, suffered by canonical methods: the
exponential increase of the tunnelling time between the metastable
states in the first-order phase transitions, when the volume of the
system is increased.  The demons act as a buffer between the
multicanonical heat bath and the spin system, allowing the spin system
to be updated with any microcanonical demon procedure, including
cluster methods.  The cluster algorithm is demonstrated with the
2-dimensional 7-state Potts model, using volumes up to $128^2$.  The
tunnelling time is found to increase as $L^{1.82}$, where $L$ is the
linear dimension of the system.

\vfill \vfill
\noindent
CERN-TH.6654/92\\
September 1992
\end{titlepage}

\section{Introduction}

If a system has a first-order phase transition, at the transition
temperature it can exist in a mixed state, where two different bulk
phases are separated by an interface.  The free energy carried by the
interface is the interface tension $\sigma$.  Because at the
transition temperature the free energy densities of the bulk phases
are equal, the free energy of the mixed state is higher than either of
the pure phases by an amount of $F_s = \sigma A$, where $A$ is the
area of the interface.  In numerical simulations using the canonical
ensemble at the transition temperature, the configurations containing
the mixed phase are suppressed by the Boltzmann factor $e^{-F_s/T}$.
When the volume of the system is increased, the suppression of the
mixed state increases exponentially with the area of the interface --
hence also the time it takes for the system to tunnel from one pure
phase to another increases exponentially.

Recently, Berg and Neuhaus \cite{Berg91} introduced a powerful new
method, the multicanonical algorithm, which avoids the exponential
slowing down by artificially enhancing the probability of the
configurations with an interface.  The tunnelling time increases only
polynomially with the increasing linear size $L$.  In this method the
individual spin updates generally depend on the total energy of the
system in a non-linear fashion.  This effectively prevents the use of
vector or parallel coding (except if one runs many lattices in
parallel) and cluster update algorithms.

In the microcanonical demon Monte Carlo approach, developed a decade
ago by M.~Creutz \cite{Creutz83}, one uses additional variables,
demons, to transfer energy around the system.  The total energy of the
system plus the demons is absolutely conserved.  However, by
periodically updating the demon energies according to the Boltzmann
distribution the method reduces to the canonical procedure.

In this work I present an algorithm which combines these two methods:
first, a system is updated {\em microcanonically} with a set of
demons, and second, the demons are refreshed with a {\em
multicanonical} heat bath.  The demons act as a buffer, isolating the
actual system from the multicanonical heat bath, thus enabling one to
choose an optimal microcanonical update step for a particular problem.
The update can be a highly vectorizable and parallelizable local
update or a cluster update, or some combination of these.  As an
example, I apply the cluster version of the algorithm to the
2-dimensional 7-state Potts model.

The Potts models have become standard tools for high-precision Monte
Carlo studies of the first-order phase transitions.  The 2-dimensional
$q$-state (2d$q$s) Potts model \cite{Potts62,Wu82} is defined by the
partition function
\be
Z(\beta) = \sum_{\{s\}} \exp[-\beta E(s)]
\ee
\be
E(s) = \sum_{(i,j)} (1-\delta(s_i,s_j)), \h\h s_i = 1,\ldots,q \, .
\ee
When $q>4$, the transition is of first order.  The infinite volume
transition temperature is $\beta_c = 1/T_c =\log(1+\sqrt{q})$.  Beside
the fact that many infinite volume quantities are exactly known,
rigorous finite-size scaling (FSS) predictions by C.~Borgs \etal
\cite{Borgs90} offer a quantitative method for studying the approach
to the asymptotic regime.  I chose the 2d7s Potts model for comparison
with the recent standard multicanonical calculation by W.~Janke \etal
\cite{Janke92}, and the canonical one by A.~Billoire \etal
\cite{Billoire92}.  The lattice sizes in this work were
$V = L^2 = 20^2$, $32^2$, $64^2$ and $128^2$; for the largest volume,
two separate simulations were performed.

This article is divided into three parts: in the first section I
discuss how the standard multicanonical approach is generalized to the
two-step demon algorithm, and how it can be used to obtain an estimate
of the density of states, and, through this, of all thermodynamical
quantities.  The next section describes the actual update algorithms:
the microcanonical cluster update and the multicanonical demon
refresh.  I present a method which enables the `slow' part of the
multicanonical update to be performed in $\propto\sqrt{V}$ steps.  The
results of the 2d7s Potts model simulations are reported in the third
section.  The tunnelling time was found to increase like
$L^{1.82(3)}$, which is better than the standard multicanonical method
($L^{2.65}$).  Where appropriate, the thermodynamical measurements of
the $128^2$ lattices fully agree with the rigorous FSS predictions of
ref.~\cite{Borgs90}, and all the measurements are consistent with the
multicanonical MC data of ref.~\cite{Janke92}.

\section{The Multicanonical Demon Algorithm}

Close to the transition temperature the canonical probability
distribution $p_\beta (E)$ develops a double-peak structure, and one
definition for the transition temperature $T^L_c=1/\beta^L_c$ itself
is the temperature when the two peaks have equal height (\fig{1}).
Following refs.~\cite{Berg91,Janke92}, the peak locations are denoted
by $E^L_1$ and $E^L_2$, and the probability density is normalized to
$p^L_\beta(E^L_1) = p^L_\beta(E^L_2) = 1$.  Denoting the minimum of
$p^L_\beta$ between the peaks by $p^L_{\min}$, the interface tension
is
\cite{Binder82}:
\be
\sigma = - \lim_{L\rightarrow\infty}
	 \frac{\log p^L_{\min}}{2L}\,.
\la{tension}
\ee
Because of the periodic boundary conditions, the configurations
corresponding to the minimum of the probability distribution have two
interfaces separated by $\sim L/2$; hence the factor 2 in
\eq\nr{tension}.  In the following the $L$-dependence of the above
quantities is mostly suppressed from the notation.


In the standard `direct' multicanonical method, one usually aims at a
roughly constant probability density in the domain between the peaks:
$p_W(E) = 1,\, E_1 \le E
\le E_2$.  This can be achieved by substituting the Boltzmann weight
with a weight function $W(E)$:
\be
p_W(E)  \propto \den_S(E)\, e^{-W(E)} \, ,
\la{mcprob}
\ee
where $\den_S(E)$ is the number of states at energy $E$.  The
requirement that the probability density is constant implies that
$W(E) \propto \log \den_S(E) = S(E)$ when $ E_1 \le E \le E_2 $; and
$W(E) = \beta_c E + \mbox{\it const.}$ otherwise.  Because $\den_S(E)$
is what we are trying to compute in the first place, we have to use an
approximate $W(E)$ instead -- obtained, for example, with finite-size
scaling, canonical simulations, or previous multicanonical
simulations.  The measured $p_W$ is then reweighted with $e^{W(E)}$ to
produce $\den_S(E)$, from which all quantities of interest can be
calculated -- also the improved $W(E)$.

Now we want to apply multicanonical ideas to a system consisting of
the original spin system and a system of demons.  Heuristically, it is
clear that the spin system in \eq\nr{mcprob} can be substituted with
any other system, also with this composite system.  Denoting the
weight function with $G$, the probability density can be written as a
joint distribution in the spin system energy $E_S$ and the demon
energy $E_D$:
\be
p_G(E_S,E_D) \propto \den_S(E_S)\den_D(E_D)\,e^{-G(E_T)} \, ,
\la{mdprob}
\ee
where $\den_S(E_S)$ and $\den_D(E_D)$ are the spin system and the
demon density of states, respectively, and $E_T=E_S+E_D$ is the total
energy.  In the most general case the weight $G$ is a function of both
$E_S$ and $E_T$, but when $E_T$ is fixed, we want $p_G$ to reduce to
the microcanonical distribution $\den_S(E_S)\den_D(E_T-E_S)$, implying
that $G$ can only be a function of $E_T$.  When $E_S$ is fixed, $p_G$
reduces to the multicanonical distribution for the demons:
$\den_D(E_D)\,e^{-G(E_S+E_D)}$.  Thus the probability distribution
\nr{mdprob} is preserved in a generic two-step process of a
microcanonical spin update and a multicanonical demon update, provided
that both of the update steps separately satisfy detailed balance.
Note that if $G(E)=\beta E$, both the demon and spin system will have
canonical distributions.

After the Monte Carlo simulation has produced a sample of the
distribution \nr{mdprob}, $\den_S$ can be solved from it.  Although
not strictly necessary, it is advantageous to use the known demon
density of states: with $N_D$ demons with discrete energy states
$0,1,\ldots$,
\be
\den_D(E_D) = \frac{(N_D - 1 + E_D)!}{(N_D-1)!\,E_D!} \, .
\la{demden}
\ee
Because \eq\nr{mdprob} is valid separately for each $E_D$, $\den_S$
can be expressed as a linear combination
\be
\den_S(E_S) \propto  \sum_{E_D} A_{E_D;E_S}\,
	\frac{p_G(E_S,E_D)}{\den_D(E_D)\,e^{-G(E_T)}}\,\, ,\h
	\sum_{E_D} A_{E_D;E_S} = 1 \, .
\la{lincomb}
\ee
The multipliers $A_{E_D;E_S}$ should be chosen to minimize the error
in $\den_S$.  For each $E_S$ separately, the uncertainty in the
(measured) $p_G$ is given by $\delta p_G = \sqrt{\bar p_G/N_{E_S}}$,
where $\bar p_G$ is the probability distribution in the limit of
infinite statistics and $N_{E_S}$ is the number of measurements with
this $E_S$.  (This is not true for the full distribution
$p_G(E_S,E_D)$ because of the correlations between successive
measurements; however, for fixed $E_S$, each measurement of $E_D$ is
completely independent, as explained below).  Minimizing the resulting
error in \eq\nr{lincomb}, we finally obtain
\be
\den_S(E_S) \propto  \frac{p_S(E_S)}
	{\sum_{E_D} \den_D(E_D)\,e^{-G(E_T)}}\, , \h
	p_{S}(E_S) \equiv \sum_{E_D} p_G(E_S,E_D) \, .
\la{denstate}
\ee
Note that the final result depends only on the spin system
distribution $p_S(E_S)$, not on the demon energy distribution.  The
distribution $p_S$ corresponds to the standard multicanonical
distribution \eq\nr{mcprob} with the weight
\be
e^{-W(E_S)} = \sum_{E_D}\den_D(E_D)\,e^{-G(E_S+E_D)} \, .
\la{weights}
\ee
Quite generally, if we want to simulate a system with a non-canonical
weight function $W(E)$, then by inverting \eq\nr{weights} we obtain
the corresponding $G(E)$. With the two-step update, the function $G$
will produce exactly the same distribution as $W$ with a direct update.

Instead of aiming at a flat distribution of $E_S$, it is now more
natural to try to `flatten' the $E_T$ distribution.  Then, the optimal
$G(E_T)$ equals to $S_T(E_T)$, the entropy of the total system.  In
the actual runs described here the initial estimate of $G(E_T)$ was
obtained (for the lattices $\le 64^2$) from short runs with $G(E_T) =
\beta E_T$; for the $128^2$ lattice, finite-size scaling was used to
scale up the function used in the $64^2$ simulation.  The $64^2$ and
$128^2$ lattices required one further refinement run to obtain the
final weight function; in the end two different weights were used for
the $128^2$ lattice.  To simplify the calculation, $E_T$ was
restricted in the range $E_T^{\min}\le E_T\le E_T^{\max}$, where the
limits were chosen such that the expectation values of $E_S$, as a
function of $E_T$, bracket the peak locations $E_1$ and $E_2$
(\fig{1}):
\be
\langle E_S\rangle(E_T^{\min}) < E_1 < E_2 <
	\langle E_S\rangle(E_T^{\max}).
\la{probrange}
\ee
The functional form of $G(E_T)$ is not crucial, as long as it is
accurate enough; here a continuous piecewise linear form was used.
The required accuracy increases with the volume of the system -- the
weight $G(E_T)$ is an extensive quantity (or rather $G(E_T)-\beta_c
E_T \propto L$), but if $G(E_T)$ is `wrong' by an amount of, say,
$\log 2$, the probability $p_G(E_T)$ will be changed by a factor of 2.

The left part of \fig{2} shows the joint distribution from the $64^2$
lattice, using $E_S$ and $E_T$ as independent variables.  As can be
seen, $E_S$ and $E_T$ follow each other very closely -- the length and
the width of the ridge behave like $ L^2$ and $L$, respectively.  The
demon energy varies only very little, meaning that the microcanonical
temperature $\partial S(E)/\partial E$ is almost constant, as it
should be in the phase coexistence region.  The right part is the same
distribution, `canonized' by reweighting it with $e^{G(E_T) - \beta_c
E_T}$, where $\beta_c$ is the transition temperature for this lattice
(table~\ref{table2}).

An interesting variation of the algorithm can be obtained by
restricting $E_T$ to a discrete set of values $E_0,\,\ldots,\,E_N$,
sufficiently dense so that the neighboring $E_S$ distributions have
large enough overlaps.  This is a microcanonical version of the
`simulated tempering' method, presented by Marinari and
Parisi\cite{Marinari92}.

\section{The Update Algorithms}

\subsection{The Microcanonical Cluster Update}

In the simulations reported here one update cycle consists of one
microcanonical spin + demon update sweep followed by an energy
measurement and a multicanonical demon update.  As the microcanonical
step I used the Swendsen-Wang variation of the microcanonical cluster
algorithm presented recently by M.~Creutz \cite{Creutz92}.  As opposed
to the standard procedure, the demons are located on the {\em links}
of the lattice, instead of being connected to the spins.  A link is
activated only if the spins have the same value at each end of it {\em
and} the demon does not carry enough energy to frustrate the link.
Clusters are grown over the activated links and each cluster is
flipped to a random spin value.  Finally, the demon energy is
increased or decreased by the amount of the corresponding link energy
decrease or increase.  On a $d$-dimensional lattice, this method
requires $d\times V$ demons.

After each update cycle the demon locations are shuffled.  If the
shuffling were not performed, one would construct exactly the same
clusters during the next update cycle -- assuming that the demon
refresh, described in the next chapter, is also skipped.  The
shuffling does not need to be perfect: here it was done with random
offsets and step lengths when picking the demons from the demon array.
The actual cluster search was performed with the Hoshen-Kopelman
cluster-finding algorithm \cite{Hoshen76}.

Note that it is also possible to perform local updates with the same
demons; the demons can be left on the links or moved to the spins --
in the latter case only half (or $1/d$) of the demons are used during
one sweep.  The probability distribution \nr{mdprob} is
unaffected, as long as the total number of demons remains the same.

\begin{table}
\center
\begin{tabular}{lrrll}
\cen{$L$} & iterations &\cen{$\tau_L$}& \cen{$\sigma$}&\\ \hline
{\n}20 & 2 500 000 &320(5)\n\n&0.0189(3)   \\
{\n}32 & 5 000 000 &821(15)\n &0.0169(2)   \\
{\n}64 & 6 000 000 &2700(81)\n&0.0147(4)   \\
$128_a$ & 9 000 000&10720(520)&0.01302(17) \\
$128_b$ & 6 000 000&10520(620)&0.01306(21) \\
\hline
\end{tabular}
\caption[1]{The tunnelling time and the interface tension from the
2d7s Potts model simulations.\la{table1}}
\end{table}

\subsection{The Multicanonical Demon Update}

The multicanonical demon refresh is greatly facilitated by the fact
that each demon is an independent degree of freedom and the demon
density of states is known.  The most straightforward way to perform
the demon update is to touch each demon with the multicanonical heat
bath: the demon $i$ is assigned a new value with the weight $\exp
[-G(E_T-E_D^{i,{\rm old}}+E^{i,{\rm new}}_D)]$.  For a continuous
demon energy this would be the best method, even though this is a
non-vectorizable process with $\sim N_D$ steps.  However, since now
the demon energy is discrete, there exists a method which enables a
major part of the demon update to be performed in $\sim\sqrt{N_D}$
(non-vectorizable) steps.  First, a new total demon energy $E^{\rm
new}_D$ is calculated with a {\em global} heat-bath update: let $x$ be
a random number from even distribution between 0 and 1; now the new
demon energy is the smallest $E_D^{\rm new}$ satisfying
\be
x \le \frac{\sum_{E'=0}^{E_D^{\rm new}} \den_D(E')e^{-G(E'+E_S)}}
	{\sum_{E''=0}^{\infty} \den_D(E'')e^{-G(E''+E_S)}} \, ,
\la{heatbath}
\ee
where $G(E_T) = \infty$, when $E_T < E_T^{\min}$ or $E_T >
E_T^{\max}$.  This guarantees that the demon energy at fixed $E_S$ is
free from autocorrelations, justifying the use of the multinomial
distribution prior to \eq\nr{denstate}.  A new demon state with energy
$E_D^{\rm new}$ can then be constructed from the old one by adding or
subtracting energy from randomly selected demons.  However, care has
to be taken to ensure proper counting of states: energy is added or
subtracted unit by unit, and the demon to be changed is chosen
according to the respective probabilities
\be
p^i_+ \equiv p^i_{E_D \rightarrow E_D+1} =  \frac{E_D^i + 1}{N_D+E_D}
\, , \h\h
p^i_- \equiv p^i_{E_D \rightarrow E_D-1} =  \frac{E_D^i}{E_D}  \, .
\la{demonadd}
\ee
To prove that \eq\nr{demonadd} is correct, it is sufficient to
show that starting from the state $E_D = 0$, $p_+$ produces with equal
probability all the
states with the same energy.  Let us
construct a specific demon state $\omega$ with an energy $E_\omega$:
by repeatedly applying $p_+$, the probability of a particular sequence
of additions $\{i\}$ leading to this state becomes
\begin{eqnarray}
 p_{\{i\}} = \prod_{\{i\}} p^i_+ & = & \frac{(N_D-1)!}{(N_D-1+E_D)!}
 \,(1!)^{N_1}\,(2!)^{N_2}\,(3!)^{N_3} \ldots \,\, , \\
 E_\omega & =  & N_1 + 2N_2 + 3N_3 + \ldots \,\, ,
\la{sequence}
\end{eqnarray}
where $N_e$ is the number of demons with energy $e$.  Because the
numbers $N_e$ are characteristic of the state $\omega$ and not of the
sequence $\{i\}$, all the sequences leading to this state have the
same probability.  Then the probability $p_\omega$ of producing the
state $\omega$ is obtained by multiplying $p_{\{i\}}$ with the number
of sequences $N^\omega_{\rm seq.}$ (= number of permutations) leading
to this state,
\be
 N^\omega_{\rm seq.} = \frac{E_\omega!}
	{(1!)^{N_1}\,(2!)^{N_2}\,(3!)^{N_3}\ldots}\, ,
\ee
giving $p_\omega = 1/n_D(E_\omega)$ [\eq\nr{demden}].  This is just
what we want -- all the states with the same energy are equally
probable, and the sum of the probabilities is 1.  It is obvious that
$p_-$ is the probability of the inverse process.  Because the old
demon state can be understood as an intermediate step in the energy
addition/subtraction sequence, we can start building the new state
directly from it.

Remember that during one update cycle only additions or only
subtractions are performed.  Because the average energy change in one
cycle is $\sim\sqrt{V}$, only as many demons need to be updated;
the demon shuffling after each cluster update takes care of proper
mixing.  By careful use of auxiliary arrays, also the integrations in
\eq\nr{heatbath} can be performed in $\sim\sqrt{V}$ steps, but
this was not fully implemented in the simulations.

An important technical question is how to implement the demon
selection according to the probabilities of \eq\nr{demonadd}.  Two
simple methods were tested: first, one can employ an appropriate
accept/reject step at a random demon location -- this is a true
$\propto\sqrt{N_D}$ method, but the poor acceptance rate (8\% for the
2d7s Potts model) makes this rather slow in practice.  In the second
method one constructs a pointer array $a$, which has an entry for the
location of each {\em unit} of the demon energy.  The array $a$ has
$E_D$ elements, of which $E_D^i$ are pointing to the demon $i$.  When
$E_D
\rightarrow E_D+1$, the demon is chosen as follows: one generates a
random integer $i$ in the range $[1, N_D+E_D]$; if $i\le N_D$, the
demon $i$ is selected, and if $i > N_D$, one chooses the demon $a_i$.
By construction, this gives just the correct probability $p^i_+$.  The
demon address is then added to the array $a$, and $E_D$ is increased.
The process is repeated until $E_D^{\rm new}$ is reached.  The
subtraction of the energy is performed analogously; only now does one
select a random pointer $a_i$, decrease the energy of the demon $a_i$,
and set $a_i = 0$.  The bottleneck in this method is the initial
generation of the pointer array, which requires $\sim{N_D}$ steps;
however, it is a vectorizable process, and since $E_D$ has to be
measured anyway, it can be performed with little overhead.  In the
tests the pointer array method was found to be 5--8 times faster than
the accept/reject procedure, so it was chosen for the actual
simulations.  On the $128^2$ lattice, the whole measurement and demon
update cycle used less than $6\%$ of the CPU time, the rest being
taken up by the cluster operations, which contain the only
non-vectorizable $\propto V$ loop.  Using a Cray X-MP, one full cycle
took $\sim 3.9\,\mu$s per spin.

\section{Results}

The performance of the algorithm is best measured by the tunnelling
time, which is defined as in ref.~\cite{Janke92}: four times the
tunnelling time, $4\tau_L$, is the average number of updates for the
system to get from $E^L_1$ to $E^L_2$ and back.  This definition gives
comparable autocorrelation time definition.  The times are listed in
table~\ref{table1}, and shown in \fig{3}.  The fit to the three
largest lattices gives $\tau_L = 1.49(17)\times L^{1.82(3)}$.  This
scales better than the standard multicanonical method \cite{Janke92},
which has $\tau_L = 0.082(17)\times L^{2.65(5)}$.  In fact, this is
even better than the {\em optimal} scaling given by the random-walk
picture: in the first-order transitions, the energy gap behaves as
$V$, but the width of the system energy distribution with fixed total
energy increases only like $\sqrt{V}$.  Assuming an ideal update
algorithm, this is also the average change in the system energy during
one sweep.  The system has to random-walk across the gap in order to
tunnel from one phase to another, and this takes $(\mbox{gap/step})^2
\sim V = L^2$ sweeps.  The discrepancy is largely due to the shift of
$E^L_1/V$ and $E^L_2/V$ (see figs.~{1} and {8}), which have a finite
size dependence $\sim 1/L$.  If we ignore these and calculate $\tau_L$
from tunnellings between the infinite volume energy density values
$e_1 = 0.4453\ldots$, $e_2 = 0.7986\ldots$, the tunnelling times of
the smaller lattices become shorter and we obtain a scaling law
$\tau_L = 0.66(7)\times L^{1.97(3)}$, which is compatible with the
random-walk limit.  However, I chose to present the former result in
order to enable the comparison with the previous calculation.  The
quality of the scaling fit is only $\chi^2 = 6.2/2$ d.o.f., and since
the `physical' correlation length is of order $\sim 30$
\cite{Billoire92}, it is quite plausible that the true scaling law
will be different.  The scaling functions are plotted in \fig{3}.

In order to compare with the canonical update algorithm, I utilized
the results of the 2d7s Potts model simulations by A.~Billoire \etal
\cite{Billoire92}.  Their work has high-statistics data from 5
different volumes between $16^2$ and $64^2$.  I made a finite-size fit
to the autocorrelation times with the heuristic function $\tau_L = a
L^\alpha\,e^{2\sigma L}$, where the interface tension $\sigma$ had the
fixed value $0.01174$ (see the next paragraph).  The parameter values
given by the fit are $a=1.01(15)$ and $\alpha=2.31(4)$, with
$\chi^2=4.2/3$ d.o.f..  (Without the exponential factor, the best
power-law fit gives $\chi^2=56/3$ d.o.f.)  The function is plotted in
\fig{3}.  The simulations in ref.~\cite{Billoire92} were performed
with the one-hit Metropolis algorithm, which has a notoriously long
autocorrelation time; this was balanced by a fast multispin coding.
With a heat-bath update the autocorrelation time could be $\sim$ 5--8
times shorter, bringing the low end of the line close to the level of
the multicanonical lines.  On the $128^2$ lattice the multicanonical
cluster method is $\sim$ 3 times faster than the standard
multicanonical, which again is $\sim$ 5--40 times faster than the
canonical method.

The interface tension was measured with the Binder method
\cite{Binder82}, \eq\nr{tension}.  The minima and maxima of the
canonical probability distributions were found by fitting a parabola
close to the extrema; the results depend on the fitting range only
very weakly, when the range is large enough.  All the error analysis
was done by jackknifing the data to 50 sets.  The measured values of
$\sigma$ are shown in \fig{4}, together with the measurements of
ref.~\cite{Janke92}.  Using a common FSS fit of the form
\be
 -\frac{1}{2L}\log p^L_{\min} = \sigma + \frac{c}{L}
\la{sigmafit}
\ee
to the three largest volumes, we obtain the result $\sigma =
0.01174(19)$ with $c=0.169(11)$.  The result agrees well with
ref.~\cite{Janke92} ($\sigma = 0.0121(5)$), but is seven standard
deviations off from the exact infinite-volume value $\sigma =
0.010396\ldots$, recently calculated by Borgs and Janke
\cite{Borgs92b}.  The cited errors are only statistical, and
the large difference between the values is obviously caused by the
violations of the FSS formula \nr{sigmafit}.  This is supported by the
missing `flat bottom' around the $p^L_{\min}$; this flat part
corresponds to the variations of the distance between the two
interfaces, and the absence of the flat part implies that the
interaction between the two interfaces is still non-negligible.

The value of $\sigma$ is at least a factor of 6 smaller than the
results of refs.~\cite{Potvin89,Kajantie89,Rummukainen91}, obtained
with an unrelated method.  In these calculations the interface was
stabilized by using different temperatures on the different sides of
the lattice, thus enforcing one side into the disordered phase and the
other into the ordered one.  The interface tension was measured as a
function of the temperature difference, and the final answer was
obtained by extrapolating the results to the transition temperature.
This method has also been applied to $N_\tau = 2$ pure gauge QCD
\cite{Kajantie90,Huang90}, and the results agree with the values
obtained with the Binder method \cite{Janke92,Grossman92}.  The
apparent failure of this method in the case of the 2d7s Potts model
is probably due to the too strong pinning effect caused by the
temperature difference.  In 2d, the amplitude of the interface
fluctuations is large, $\sim\sqrt{L/\sigma T}$.  These fluctuations
are strongly suppressed by the temperature difference, which should
then be very small; however, in that case the two-phase
configuration would be lost, unless extremely large volumes are used.


\begin{table}
\center
\begin{tabular}{lllll}
\cen{$L$} & \cen{$\beta$(equal height)}&\cen{$\beta(C_{\max})$}&
	\cen{$\beta(B_{\min}$)}&\cen{$\beta$(equal weight)}\\ \hline
{\n}20      &1.28474(13) &1.28443(12)&1.28444(13) &1.2939(3) \\
{\n}32      &1.28976(7)  &1.28953(7) &1.28776(7)  &1.29379(7) \\
{\n}64      &1.29241(4)  &1.29235(3) &1.29194(3)  &1.29360(4) \\
$128_a$ &1.293251(15)&1.293234(15)&1.293140(15)&1.293567(16)\\
$128_b$ &1.293242(19)&1.293227(19)&1.293133(19)&1.293560(19)\\
\hline
\end{tabular}
\caption[1]{Measured pseudotransition temperatures.\la{table2}}
\end{table}

Finally, let us compare transition temperature measurements to the
exact finite-size expansions \cite{Borgs90,Lee91}.  Common definitions
for the pseudotransition temperature are the locations of the maximum
of the heat capacity $C = \beta^2/V\, (\lg E^2\rg -\lg E\rg^2)$ and
the minimum of the Binder parameter $V_L =\frac{1}{3}(1-\lg E'^4\rg
/\lg E'^2\rg^2)$, where $E' = E-2V$ in order to comply with the
definition of energy used in
\cite{Borgs90,Janke92}.  The differences of the known infinite-volume
transition temperature and the measured values are plotted in
\fig{5}; also shown are the exact lowest order FSS corrections.
On the $128^2$ lattices the FSS correction is within the error bars of
the measured values, whereas the $64^2$ lattice is still off by $\sim
2\sigma$.  Figure {6} shows the behaviour of $V_L$ as a function of
$\beta$.

Still another way to find the transition temperature is the `equal
weight' method \cite{Borgs92}, where $\beta_w^L$ is defined as the
temperature when the relative probabilistic weights of the ordered and
disordered states are $q=7$ and 1, respectively:
\be
 q = W_O/W_D = \sum_{E<E'} p_{\beta_w^L}(E)/\sum_{E\ge E'} p_{\beta_w^L}(E),
\la{eqweight}
\ee
where $E'$ is the energy at the minimum of $p_{\beta}$ at the
temperature when the two peaks have equal height.  The measurements of
$\beta_c$ are shown in \fig{7}.  The FSS corrections in this case are
only exponential.  An FSS ansatz of the form $\beta_w^L = \beta_w +
a\,e^{-bL}$ was fitted to the data, with the results $a=0.0012(12)$,
$b=0.05(3)$, and $\beta_w = 1.293562(14)$.  The fit gives exactly the
correct $\beta_c$; however, $\beta_w$ is almost completely determined
by the $128^2$ data, and the effect of the values of $a$ and $b$ to
the value of $\beta_w$ is negligible.  The various transition
temperature measurements are listed in table~\ref{table2}.

The locations of the maxima of the canonical probability distribution
$p_\beta$ are shown in \fig{8}.  The FSS fits to the three largest
volumes give the infinite-volume result $e_{1,\infty} = 0.4421(24)$
and $e_{2,\infty} = 0.7981(17)$, which agree fairly well with the
exact values $0.44539\ldots$ and $0.79867\ldots$.  The latent heat is
the difference of these two values: $\Delta e = 0.3559(29)$ (exact
$0.35327\ldots$).

Even though the difference between the heat capacity of the disordered
phase ($C_{\rm D})$ and of the ordered phase ($C_{\rm O}$) is exactly
known, the actual pure phase values are not.  An estimate can be
obtained by employing the FSS relation of refs.~\cite{Lee91,Janke92}:
$C_{\rm O} = C_{L,\max} - V (\Delta s/2)^2 + 0.0038\ldots + {\cal
O}(V^{-1})$, where $\Delta s = \beta_c \Delta E/V$ is the entropy
difference between the two phases.  The result is shown in \fig{9};
the FSS fit yields an estimate of $C_{\rm O} = 44.4 \pm 2.2$.  Because
of the subtraction of two terms of order $V$, the errors grow rapidly
when the volume is increased.  Again, this result is consistent with
that of ref.~\cite{Janke92} $47.5 \pm 2.5$.

\section{Conclusions}

I have presented a new hybrid-like algorithm, which combines a
microcanonical spin system update with demons, and a multicanonical
demon update.  Like the direct multicanonical method, this algorithm
does not suffer from the exponential slowing down in the first-order
phase transitions.  In the 2d7s Potts model simulations, the
tunnelling time was found to increase like $\tau_L \sim L^{1.82(3)}$
with lattices up to $L^2 = 128^2$.  Where appropriate, the
measurements were compared with the analytical finite-size scaling
formulas by Borgs \etal \cite{Borgs90}; within the statistical errors,
the $128^2$ lattice was found to be in complete agreement with the
order $1/V$ FSS predictions.  Also, all results are fully compatible
with those of ref.~\cite{Janke92}.  However, the common FSS ansatz for
the interface tension, $\sigma_L = \sigma_\infty + c/L$, fails to
produce the correct infinite-volume value.  This is probably due to
the interactions between two interfaces, which are ignored by the FSS
ansatz.  Clearly, a better FSS function is needed.

The main advantage of the multicanonical demon algorithm is that it
offers various possibilities in choosing the algorithm.  In addition
to simply using either local or cluster updates, one can also adjust
the number of the demons and the number of the microcanonical updates
before each multicanonical step.  For example, if one has a very fast
local microcanonical algorithm, it might be preferable to interleave
many microcanonical updates for each demon refresh, and to use a large
number of demons ($N_D > V$) in order to allow large fluctuations in
the system energy during the microcanonical phase -- the demon refresh
can still be performed in $N_D\times\mbox{fast} +
\sqrt{N_D}\times\mbox{slow}$ operations.  The method can also be
generalized to magnetic transitions by using demons carrying
magnetization; in this case the cluster algorithm cannot be used.

\section*{Acknowledgments}

I am grateful to Leo K\"arkk\"ainen, A.~Irb\"ack, S.~Gupta and
W.~Janke for helpful discussions.  The simulations were performed with
Sun ELC workstations and Cray X-MPs at CERN and at the Centre for
Scientific Computing, Helsinki.

\pagebreak

\subsection*{Figure captions}

\noindent
Fig.~1.
The canonical probability distributions $p_{\beta_c}(E)$
obtained from the multicanonical distributions with
\eq\nr{denstate}.\vspace{3mm}

\noindent
Fig.~2.
The multicanonical joint probability distribution
$p_G(E_S,E_T)$, measured from the $64^2$ 7s Potts model simulation
(left) and the canonical distribution corresponding to $\beta=1.29241$
(right). The canonical distribution is obtained from the
multicanonical one by reweighting it with $e^{G(E_T)-\beta
E_T}$.\vspace{3mm}

\noindent
Fig.~3.
The tunnelling time as a function of the lattice size $L$
on a log-log scale with scaling laws.  Solid line (a): this work,
$\tau_L \propto L^{1.82}$; dash-dotted line (b): multicanonical
simulation by Janke \etal \cite{Janke92}, $L^{2.65}$; dashed line (c):
canonical simulation by Billoire \etal \cite{Billoire92},
$L^{2.3}\,e^{2\sigma L}$.
\vspace{3mm}

\noindent
Fig.~4.
The interface tension as a function of the lattice size.
The black circles correspond to this work, and white squares to
measurements by Janke \etal \cite{Janke92}.  The line is a fit to 3
largest volumes, and the white circle is the infinite volume
extrapolation.  The arrow is the exact infinite volume result
\cite{Borgs92b}.
\vspace{3mm}

\noindent
Fig.~5.
The difference between the infinite volume transition temperature
$\beta_c^\infty$ and the measured transition temperature $\beta_c^L$,
plotted on a log-log scale.  The black circles are the measured
locations of the maxima of the heat capacity ($C_{\max}$) and the
white circles the locations of the minima of the Binder parameter
($V_{L,\min}$).  The solid line and the dashed line are the exact
order $1/V$ expansions.  For clarity, the data of the second $128^2$
lattice has been shifted slightly to the right.
\vspace{3mm}

\noindent
Fig.~6.
The Binder parameter $V_L$.  The dashed line shows the exactly known
infinite volume limit.
\vspace{3mm}

\noindent
Fig.~7.
The transition temperature determined by the equal weight
method, and the FSS fit to the data.  The dashed line is the exact
infinite-volume $\beta_c$.
\vspace{3mm}

\noindent
Fig.~8.
The locations of the maxima of the canonical probability distribution
$p^L_{\beta}$, together with the linear fit to three largest
volumes.
\vspace{3mm}

\noindent
Fig.~9.
The maximum of the heat capacity with the leading finite size
correction subtracted.

\hfill

\end{document}